\documentclass[12pt,preprint]{aastex}
\renewcommand{\cite}{\citealp}

\newcommand{\realfigure}[3]{ 
             \hbox{~} \centerline{\includegraphics[width=3.2in]{#1}}
             \figcaption{#2 \label{#3}} \vspace{0.05in}\centerline{}}

\shorttitle{The CVn\,II dSph}
\shortauthors{Greco et al.}

\begin{document}

\title{On the 
newly discovered 
Canes Venatici\,II dSph galaxy\altaffilmark{1}}
\author{Claudia Greco\altaffilmark{2,3},  
Massimo Dall'Ora\altaffilmark{4},
Gisella Clementini\altaffilmark{2},
Vincenzo Ripepi\altaffilmark{4},
Luca Di Fabrizio\altaffilmark{5},
Karen Kinemuchi\altaffilmark{6},
Marcella Marconi\altaffilmark{4},
Ilaria Musella\altaffilmark{4},
Horace A. Smith\altaffilmark{7},
Christopher T. Rodgers\altaffilmark{8},
Charles Kuehn\altaffilmark{7},
Timothy C. Beers\altaffilmark{7,9},
M\'arcio Catelan\altaffilmark{10}, 
and Barton J. Pritzl, \altaffilmark{11}}
\altaffiltext{2}{INAF, Osservatorio Astronomico di
Bologna, Bologna, Italy; gisella.clementini@oabo.inaf.it}
\altaffiltext{3}{Current address: Observatoire de Geneve, Sauverny, Switzerland; Claudia.Greco@obs.unige.ch}
\altaffiltext{4}{INAF, Osservatorio Astronomico di Capodimonte, Napoli, Italy,
dallora@na.astro.it, ripepi@na.astro.it, marcella@na.astro.it, ilaria@na.astro.it}
\altaffiltext{5}{INAF, Centro Galileo Galilei \& Telescopio Nazionale Galileo, S.
Cruz de La Palma, Spain; difabrizio@tng.iac.es}
\altaffiltext{6}{Universidad de Concepci\'on, Departamento de
F\'{\i}sica, Concepci\'on, Chile, and University of Florida, Department of
Astronomy, Gainesville, FL 32611-2055, USA;
kkinemuchi@astro-udec.cl}
\altaffiltext{7}{Department of Physics and Astronomy, Michigan State University, East Lansing, 
MI 48824-2320, USA; smith@pa.msu.edu, kuehncha@pa.msu.edu, beers@pa.msu.edu}
\altaffiltext{8}{University of Wyoming, Department of Physics \& Astronomy,  
Laramie, WY 82071, USA; crodgers@uwyo.edu}
\altaffiltext{9}{Joint Institute for Nuclear Astrophysics, Michigan State
University, East Lansing, MI 48824, USA}
\altaffiltext{10}{Pontificia Universidad Cat$\rm{\acute{o}}$lica de Chile,
Departamento de Astronom\'{\i}a y Astrof\'{\i}sica, Santiago, Chile; mcatelan@astro.puc.cl}
\altaffiltext{11}{Department of Physics and Astronomy, University of Wisconsin Oshkosh, Oshkosh, WI 54901, USA; pritzlb@uwosh.edu}

\altaffiltext{1}{Based on data collected at the 4.2m William Herschel Telescope at 
Roche de los Muchachos, Canary Islands, Spain, and at the 2.3m
telescope at the Wyoming Infrared Observatory (WIRO) at Mt. Jelm, Wyoming, USA.}

\begin{abstract}
We report on the detection of variable stars in the Canes Venatici\,II (CVn\,II)
dwarf spheroidal galaxy, a new satellite of the Milky Way recently discovered by the Sloan Digital Sky Survey.  
We also present a $V, \bv$ color-magnitude diagram that reaches $V \sim 25.5$ mag, 
showing the galaxy's main sequence  
turn off at $V \sim 24.5$ mag and revealing several candidate blue straggler stars.
Two RR Lyrae stars have been identified within the half-light radius of CVn\,II, a 
fundamental-mode variable (RRab) with period $P_{ab}$ = 0.743 days, and a first-overtone (RRc) RR Lyrae star
with $P_{c}$ = 0.358 days. The rather long periods of these variables along with their position on the 
period-amplitude diagram support an Oosterhoff type II classification for CVn\,II. 
The average apparent magnitude of the RR Lyrae stars, 
$\langle  V\rangle = 21.48 \pm 0.02$ mag, 
is used to obtain a precision distance modulus of 
$\mu_0 =21.02 \pm 0.06$ mag and a corresponding distance
of 160$^{+4}_{-5}$ kpc, for an 
adopted reddening $E(\bv)=0.015$ mag.

\end{abstract}


\section{Introduction}
Dwarf spheroidal (dSph) galaxies represent the most numerous class of
objects in the Local Group (LG).  They are also the most dark-matter
dominated ones, thus making them prime candidates to host a large fraction of 
the 
mass in the Universe \citep{mateo98} and the most suitable candidates among which to
search for the ``building blocks'' that may have contributed parts of
the halos of larger galaxies.
In the last couple of years, ten new Milky Way (MW) satellites have
been discovered by the Sloan Digital Sky Survey (SDSS) \citep{yo00},
namely: Ursa Major\,I and II (UMa\,II), Canes Venatici\,I (CVn\,I) and II (CVn\,II), 
Bootes\,I and II,
Leo\,IV,
Hercules, and Leo\,T
(\citealt{will05,zu06a,zu06b,be06,be07,wal07,gr06,irw07}).  Along with
the 10 previously known MW dSph companions (Draco, Ursa
Minor, Fornax, Carina, Sculptor, Leo\,I, Leo\,II, Sextans; \citealt{mateo98}), and the two
accreting systems, Sagittarius (Ibata, Gilmore, \& Irwin 1995) and Canis Major (\citealt{martin04}), 
these new discoveries bring to twenty the number of dSph's surrounding
the MW.
The new dwarfs have half-light radii resembling those of the classical
dSph's, however, they all have an effective surface brightness $\mu_V
\gtrsim 28$ mag arcsec$^{-2}$ (\citealt{be07}), hence are
fainter than the previously known LG dSph's.
Their shapes are also quite irregular and there seems to be a
correlation between irregularity and distance, as the closest ones
(namely 
Bootes\,I, UMa\,II and Coma) appear to be more distorted,
as if they were torqued by tidal interaction with the MW
(\citealt{be06,be07,zu06b}).  All the new systems appear also to be
rather metal-poor and to host a dominant old stellar population
(\citealt{zu06a,will05,be06,be07,martin07,sg07}).

CVn\,II (R.A. = 12$^{\rm h}$57$^{\rm m}$10$^{\rm s}$, 
DEC = 34$^{\circ}$19$^{\prime}$15$^{\prime \prime}$, J2000.0;
$\ell = 113.6^{\circ}$, $b= 82.7^{\circ}$) is one of the faintest of
the newly discovered SDSS galaxies, with $M_{V} = -4.8 \pm 0.6$ mag
and surface brightness $\sim$ 29.5 mag arcsecond$^{-2}$ (\citealt{be07}).
It is a low-mass ($[2.4 \pm 1.1] \times 10^6 M_\odot$; 
\citealt{sg07}),
compact dSph with a half-light radius $r_h$= $3.0^{\prime} \pm 0.5
^{\prime}$ (\citealt{be07}), corresponding to 132 $\pm$ 15 pc at a
distance $d = 151^{+15}_{-13}$ kpc.  The $i, g-i$ color-magnitude
diagram (CMD) by \citet{be07} shows that CVn\,II has a well-defined
and narrow red giant branch and a horizontal branch (HB) that extends
to the blue, through the RR Lyrae instability strip.  \citet{sg07}
obtained spectra for 24 bright stars in CVn\,II, from which they
derived a velocity dispersion of 4.6 $\pm$ 1.7 km sec$^{-1}$ and an
average metal abundance [Fe/H]=$-2.31 \pm 0.12$ dex with a dispersion of
$\sigma_{\rm [Fe/H]}$=0.47 dex using the \citet{ru97} technique, which
provides metal abundances consistent with the \citet{cg97} metallicity
scale.

In this {\em Letter} 
we present a first CMD of the CVn\,II dSph galaxy in the
$B,V$ bands of the Johnson-Cousins photometric system, extending 
down to $V \sim 25.5$ mag and revealing the galaxy's main sequence 
turn off at $V \sim 24.5$ mag. We also provide light curves for two RR Lyrae stars
we have identified in CVn\,II.

\section{Observations and Data Reduction}

Time-series $B,V,I$ photometry of the CVn\,II dSph galaxy was
collected in 2007, May 10-12, using the Prime Focus Imaging Camera
(PFIP) of the 4.2m William Herschel Telescope (WHT), and the
WIRO-Prime, the prime focus CCD camera (\citealt{p02}) of the 2.3m
Wyoming Infrared Observatory telescope (WIRO).  The WHT and WIRO
observations cover fields of view (FOVs) measuring approximately 
$16.2 \times 16.2$ arcmin$^2$ and $17.8 \times 17.8$ arcmin$^2$ in size, 
respectively, and allow us both to
completely map the galaxy and to infer the contamination by field
stars and background galaxies using an external area devoid of CVn\,II
stars.

We obtained 30 $V$, 30 $B$ and 15 $I$ frames in total,
corresponding to total exposure times of about 5\,h, 5\,h, and 2.5\,h
in $V$, $B$, and $I$, respectively. 
In this {\em Letter}  we present results from the analysis of the 
$B$ and $V$ data. 

Images were pre-reduced following standard procedures (bias
subtraction and flat-field correction) with IRAF. 
We then performed
PSF photometry with the
DAOPHOT\,IV/ALLSTAR/ALLFRAME packages (\citealt{st87,st94}). 
The absolute photometric calibration was performed using standard
stars in Landolt's (1992) field PG0918 observed at the WHT
during the night of 2007, May 12. To derive individual calibration
equations, all the PG0918 standards were observed 
in each of the two chips composing the PFIP camera.  The atmospheric
extinction coefficients were calculated directly from the time-series
observations of CVn\,II taken at different airmasses during the same
night.  A total number of 7 standard stars covering the
color interval $-0.3 \lesssim \bv \lesssim 1.3$ mag were used to 
derive the calibration
equations.
The resulting scatter was less than 0.01 mag
in both 
filters and for both chips of the WHT mosaic.
Typical errors
at the level of the CVn\,II HB ($V \sim 21.5$ mag) for the combined
photometry of non-variable stars are 
$(\sigma_{V},\,\sigma_{B}) = (0.005,\,0.01)$ mag and 
$(\sigma_{V},\,\sigma_{B}) = (0.01,\,0.015)$ mag 
for the WHT and WIRO datasets, respectively.

\section[]{Identification of the variable stars} 

Variable stars were identified from both the $B$ and the $V$ time series.
First we calculated the Fourier transform (in the \citealt{sc96}
formulation) for each star in the photometric catalog with at least 12
epochs, then we averaged this transform to estimate the noise and
calculated the signal-to-noise ratios. The results in $V$ and $B$ were 
cross-correlated to eliminate spurious detections.
We then checked all the stars with high S/N, and in particular all the stars 
around the HB. Since CVn\,II seems to host several blue straggler 
stars (BSS) (see \S4)
we also checked whether some of the stars in the BSS region might be
variables of SX Phoenicis type. Possible candidates were found; however, no
conclusive results were reached because of the severe aliasing problems
caused by the data windowing. 
Study of the light curves and period derivation were carried out using
GRaTiS (Graphical Analyzer of Time Series), which is proprietary software
developed at the Bologna Observatory (see
\citealt{df99,clementini00}).  We confirmed the variability and
obtained reliable periods and light curves for 2 RR Lyrae stars: 1
fundamental-mode (RRab) variable with period $P=0.743$ days, and 1 first
overtone (RRc) star with period $P=0.358$ days. Both stars lie within
the half-light radius of the galaxy and fall on the HB of the CVn\,II
CMD, thus strongly supporting 
their membership to the galaxy.  
Identification and properties of the confirmed variable
stars are summarized in Table~1.

In the MW, GCs that contain significant numbers of RR Lyrae stars have
the mean period of the fundamental-mode pulsators ($\langle
P_{ab}\rangle$) either of about 0.55 days or of about 0.65 days, and
separate into the so-called Oosterhoff I (Oo\,I) and Oosterhoff II
(Oo\,II) types (\citealt{oo39}).
Extragalactic globular clusters and field RR Lyrae stars in dSph
galaxies instead generally have the $\langle P_{ab}\rangle$
intermediate between the two types (\citealt{ca04,ca05}).
The rather long periods of the RR Lyrae stars indicate that CVn\,II is
an Oosterhoff type II system.  So far, only two dSph galaxies of pure
Oosterhoff type II had been known, namely Ursa Minor (UMi) among the
traditional companions of the MW, and 
Bootes\,I
(\citealt{dall06,s06}) among the newly discovered SDSS dSph's.
In terms of RR Lyrae star properties, CVn\,II thus resembles UMi and
Bootes\,I, and differs instead from CVn\,I, the brightest of the
the SDSS dSph's, which has an Oosterhoff-intermediate type
(\citealt{kue07}).

In Figure~\ref{f:fig2} we plot the CVn\,II RR Lyrae stars on the $V$
and $B$ period-amplitude diagrams of the 
Bootes\,I dSph galaxy,
using data from \citet{dall06} and
\citet{s06}.
The position of the CVn\,II variables in Figure~\ref{f:fig2} confirms
their similarity to the 
Bootes\,I RR Lyrae stars and supports the
classification of CVn\,II as an Oosterhoff type II system.

\section{The CMD and the galaxy structure}
The $V, \bv$ CMDs of the CVn\,II dSph are shown in
Figure~\ref{f:fig3}, where we have plotted
objects in the whole 16.2 $\times$ 16.2 arcmin$^2$ field covered by the 
 WHT observations in panel {\it a}; only objects 
 within the galaxy's half-light radius ($r = 3.0^{\prime}$) in panel {\it b};  
sources located 
in an annulus at $3.0^{\prime} < r < 4.2^{\prime}$ 
in panel {\it c}; and,
in panel {\it d}, sources located in an external annular region at 
$7.4^{\prime} < r < 8.0^{\prime}$. 
The three regions cover areas in the ratio 1:1:1.
Only stars with $\sigma_{V}$, $\sigma_{B} \leq 0.10$ mag , $\chi \leq$
2 are plotted in the figure. The solid and dashed lines 
are, respectively, the mean ridge lines of the
Galactic GCs M15 (NGC\,7078)
and M3 (NGC\,5272), 
drawn from the CMDs by \citet{dh93} for M15 and \citet{buo94} for M3,
shifted in magnitude and color
to match the CVn\,II horizontal and red giant branches. 
Adopting for
the reddening of M15 $E(\bv)=0.10\pm$0.01 mag (\citealt{dh93}), and
for M3 $E(\bv)=0.01 \pm$0.01 mag (\citealt{ha96}), the color shifts
required to match the CVn\,II HB to those of M15 and M3 thus imply a
reddening $E(\bv)=0.015 \pm$0.010 mag for CVn\,II. This is in
agreement with the 0.014 $\pm$ 0.026 mag value derived for the galaxy
from the \citet*{sc98} maps. 
The CVn\,II CMD reaches $V \sim 25.5$ mag, 
and when considering the whole field of view of the WHT observations
(Fig. 3{\it a}) appears to be heavily contaminated by field objects
at every magnitude level. 
The HB of CVn\,II shows up quite clearly; however, the galaxy's red giant
branch (RGB) is barely discernible from contaminating stars belonging
to the MW halo and disk.  
Moreover, inspection of the frames reveals
that CVn\,II is surrounded and partially embedded into clusters of
background galaxies. In an attempt to separate stars from galaxies, we
ran Source Extractor (SExtractor, \citealt{ber96}) on the WHT data.
The morphological parameters of the detected sources allowed us to
discriminate between point sources and extended objects for magnitudes
brighter than $V \sim 24.4$ mag.  The confirmed bona-fide stars are
marked in blue and green in Figure~\ref{f:fig3}.  Figure 3{\it b} shows that the
contamination by field sources can be significantly reduced if we
consider only stars within the galaxy's half-light radius. Indeed, the
features of the galaxy's CMD show up much more clearly in Figure 3{\it b}
and can be traced down to the main sequence turn off at $V\sim 24.5$ mag,
where several stars are located in the BSS region (green dots in 
Fig. 3{\it b}), and are likely BSSs of CVn\,II. The HB and RGB of
CVn\,II are very well reproduced by the CMD of the Galactic globular
cluster M15, implying that CVn\,II has an old and metal-poor stellar
population with metal abundance comparable to that of M15: ${\rm
  [Fe/H]}=-2.15 \pm 0.08$ dex or $-2.12 \pm 0.01$ dex on the \citet{zw84}
and \citet{cg97} scales, respectively. On the other hand, to match the M3
RGB would require a negative reddening for the cluster, 
and makes any metallicity spread as large as 0.5 dex or larger in CVn\,II  unlikely, since the
metallicity of M3 is ${\rm [Fe/H]}=-1.66 \pm 0.06$ dex or $-1.34 \pm 0.02$
dex on the \citet{zw84} and the \citet{cg97} scales, respectively. Although 
SExtractor does not provide a reliable discrimination between stars
and galaxies fainter than $V$=24.4 mag, Figure 3{\it b} suggests
that many of the faint sources falling close to the ridge line of the
M15 mean sequence are very likely main sequence stars of the CVn\,II
dSph. 
Stars belonging to CVn\,II are still seen in Figure 3{\it c};
however, they appear to be rather few in number, thus indicating that the
vast majority of the CVn\,II stars are confined within $r <
3.0^{\prime}$ from the galaxy center. Finally, Figure 3{\it d} is a
control field devoid of CVn\,II stars, and provides an indication of
the degree of contamination by field stars and background galaxies in
the region of the CVn\,II dSph.

We have used the mean ridge line of M15 as a reference to locate the stars most likely 
belonging to the CVn\,II galaxy in the CMDs of Figure~\ref{f:fig3}. In 
Figure~\ref{f:fig4} we show the X,Y map of the sources that in the CMDs 
lie within $\pm$0.1 mag from the mean ridge line of M15 
(stars are shown in blue, 
whereas objects classified by SExtractor as non-stellar or non-classified 
are shown in black) 
and of the stars
with $V < 24.4$ mag in the region of the BSSs (green dots).
Circles locate the 3 different areas corresponding to the CMDs
in  Figs.~3{\it b,c,d}. 
The bulk of the stars most likely to be members of the CVn\,II dSph is located within 
the inner circle of radius $r= 3.0$ arcmin, which corresponds to the galaxy's half-light radius.
These stars outline a mainly circular structure perhaps slightly elongated in 
the South-West direction. 
 
The strong similarity of the CMD 
to that of a simple
population system like M15, the roughly spherical distribution of stars
around the galaxy center (see Figure~\ref{f:fig4}), and the
compactness and relatively small size might suggest that CVn\,II
is indeed a metal poor globular cluster, rather than an actual dSph
galaxy. 
However, at a
distance of 151 kpc (\citealt{be07}), the half-light radius of CVn\,II
corresponds to 132 pc, hence about 5-6 times larger than observed
for NGC\,2419, the largest of the Galactic GCs according to \citet{bell07}.
Such a large size is clearly inconsistent with the tight relation between size 
and galactocentric distance derived by van den Bergh (1995) for Galactic GCs, 
thus strongly disfavoring its GC classification.

\section[]{An Improved Distance to the CVn\,II galaxy}

The average apparent luminosity of the CVn\,II
RR~ Lyrae stars is $\langle
V\rangle =21.48 \pm 0.02$ mag (average on 2 stars).
Assuming $M_V$=0.59$\pm$0.03 mag 
for the absolute luminosity of the RR Lyrae stars at [Fe/H]=$-1.5$ dex (\citealt{cc03}), 
$\Delta M_V/\Delta [Fe/H]$=0.214 ($\pm$ 0.047) mag/dex for the slope of the 
luminosity-metallicity relation (\citealt{cg03,gratton04}), 
 $E(B-V)$=0.015 mag, and [Fe/H]=$-2.31$ dex (\citealt{sg07}), 
 the distance modulus of CVn\,II is 21.02 $\pm$ 0.06 mag, which corresponds to 
 a distance $d$=160$^{+4}_{-5}~\rm{kpc}$. Errors include the standard
 deviation of the mean, and the uncertainties in the photometry, reddening, and
 RR Lyrae absolute magnitude. This new, precise distance estimate agrees, within the uncertainties, 
 with the distance of $151^{+15}_{-13}~\rm{kpc}$ 
found by \citet{be07}.

\section[]{Summary and conclusions}

We have identified and obtained $B,V$ light curves for 2 RR~ Lyrae stars 
(1 RRc and 1 RRab) in the newly discovered CVn\,II dSph galaxy
(\citealt{be07}). The behavior of these two pulsators suggests 
an Oosterhoff II classification of CVn\,II. 
From the average luminosity of the RR Lyrae stars, 
the galaxy's distance modulus
is  $\mu_0$=21.02 $\pm$ 0.06 mag ($d=160^{+4}_{-5}~\rm{kpc}$). 

\bigskip

We warmly thank Luciana Federici for providing the mean ridge lines of M3. 
Financial support for this study was provided 
by PRIN-INAF 2006 (PI G. Clementini). 
HAS thanks 
the U.S. NSF for support under grant AST0607249. 
MC is supported by Proyecto 
Fondecyt \#1071002.  


\clearpage

  \begin{table*}
  \small
      \caption[]{Identification and properties of the RR Lyrae stars in the CVn\,II dSph
      galaxy}
         \label{t:bootes_var}
     $$
         \begin{array}{lcclllccrr}
	    \hline
            \hline
           \noalign{\smallskip}
           {\rm Name} &  {\rm \alpha } & {\rm \delta} &  {\rm Type} &~~~~P & 
	    ~~~{\rm Epoch (max)}  & \langle V\rangle  & \langle B\rangle  & A_V~~ & A_B~~ \\
            ~~ & {\rm (2000)}& {\rm (2000)}& & ~{\rm (days)}& ($-$2450000) &(a) & (a) 
	     & {\rm (mag)} &  {\rm (mag)}\\
            \noalign{\smallskip}
            \hline
            \noalign{\smallskip}
	    
{\rm ~V1} & 12:57:01.6 & 34:19:33.4 & {\rm RRc }     &0.358 &4232.854   & 21.49 & 21.74 &0.678&0.832\\ 
{\rm ~V2} & 12:57:11.8 & 34:16:52.9 & {\rm RRab }    &0.743 &4231.504   & 21.46 & 21.77 &0.707&0.952\\ 
\hline
            \end{array}
	    $$
{\small $^{\mathrm{a}}$ ${\rm \langle V\rangle}$ and ${\rm \langle B\rangle}$ values are 
intensity-weighted mean magnitudes.}\\
\end{table*}

\clearpage

\realfigure{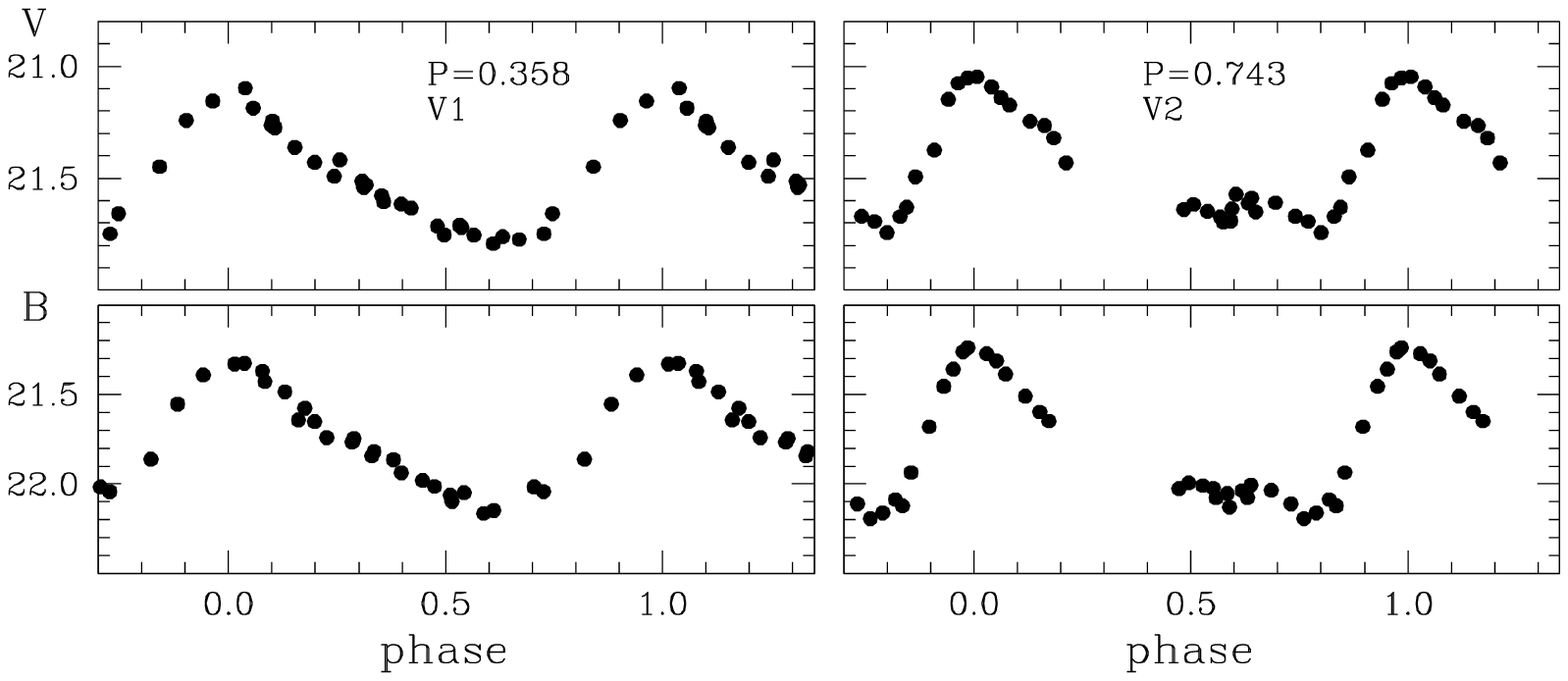} {$V$ and $B$ light curves of the two RR Lyrae
  stars discovered in CVn\,II.  {\it Left panels}: {\it c-}type RR Lyrae star; 
  {\it right panels}: {\it ab-}type RR Lyrae star.}
{f:fig1} 

\clearpage

\realfigure{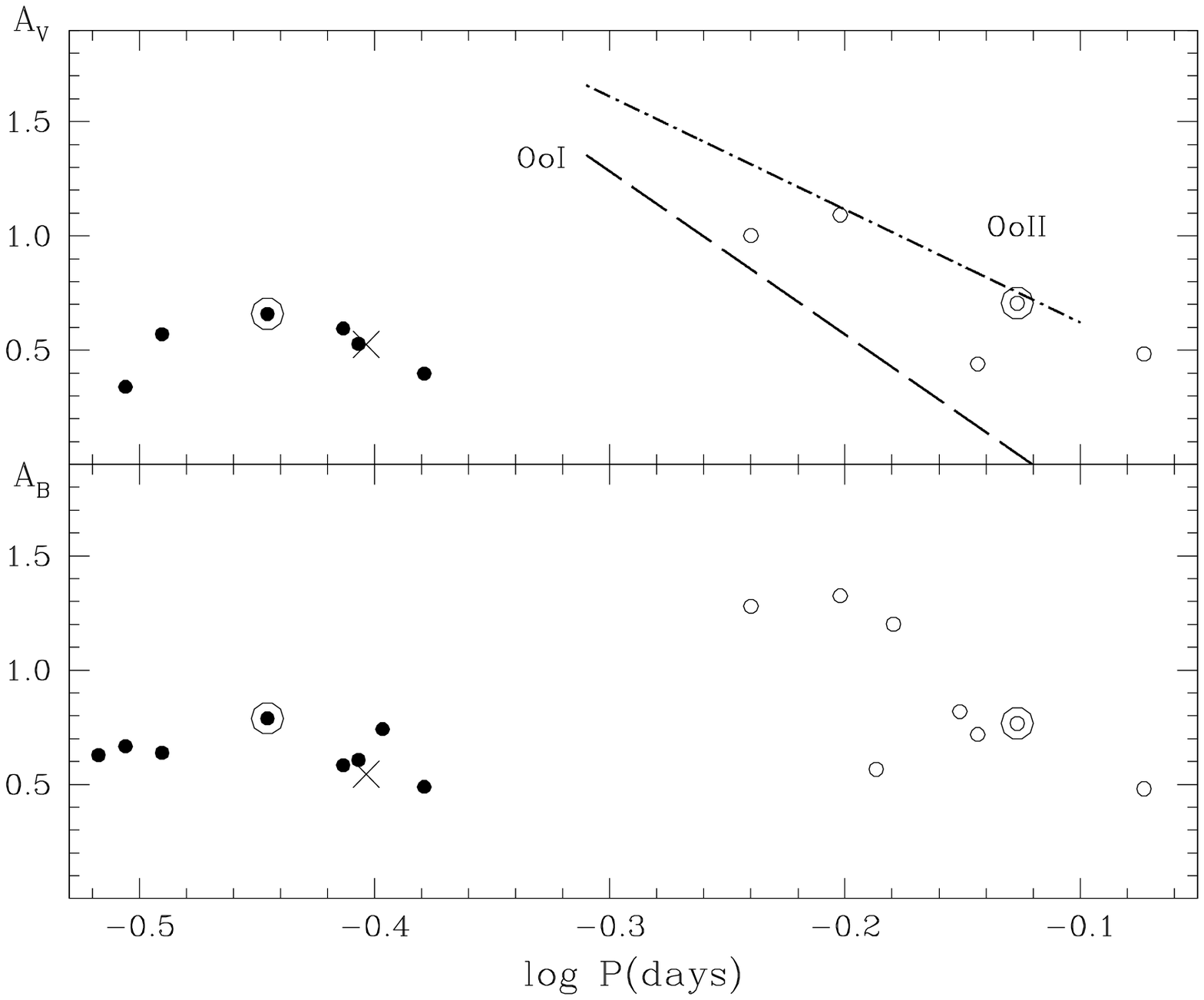} {Position of the CVn\,II RR Lyrae stars (double
  circles) on the $V$ and $B$ period-amplitude diagrams of the
  Bootes\,I dSph variables.  Open and filled circles are
  fundamental-mode and first-overtone pulsators, respectively. The
  cross is a double-mode variable. The two lines show the positions of
  Oo\,I and Oo\,II Galactic GCs according to \citet{cr00}.}{f:fig2}

\clearpage

\begin{figure*}
\includegraphics[scale=.90]{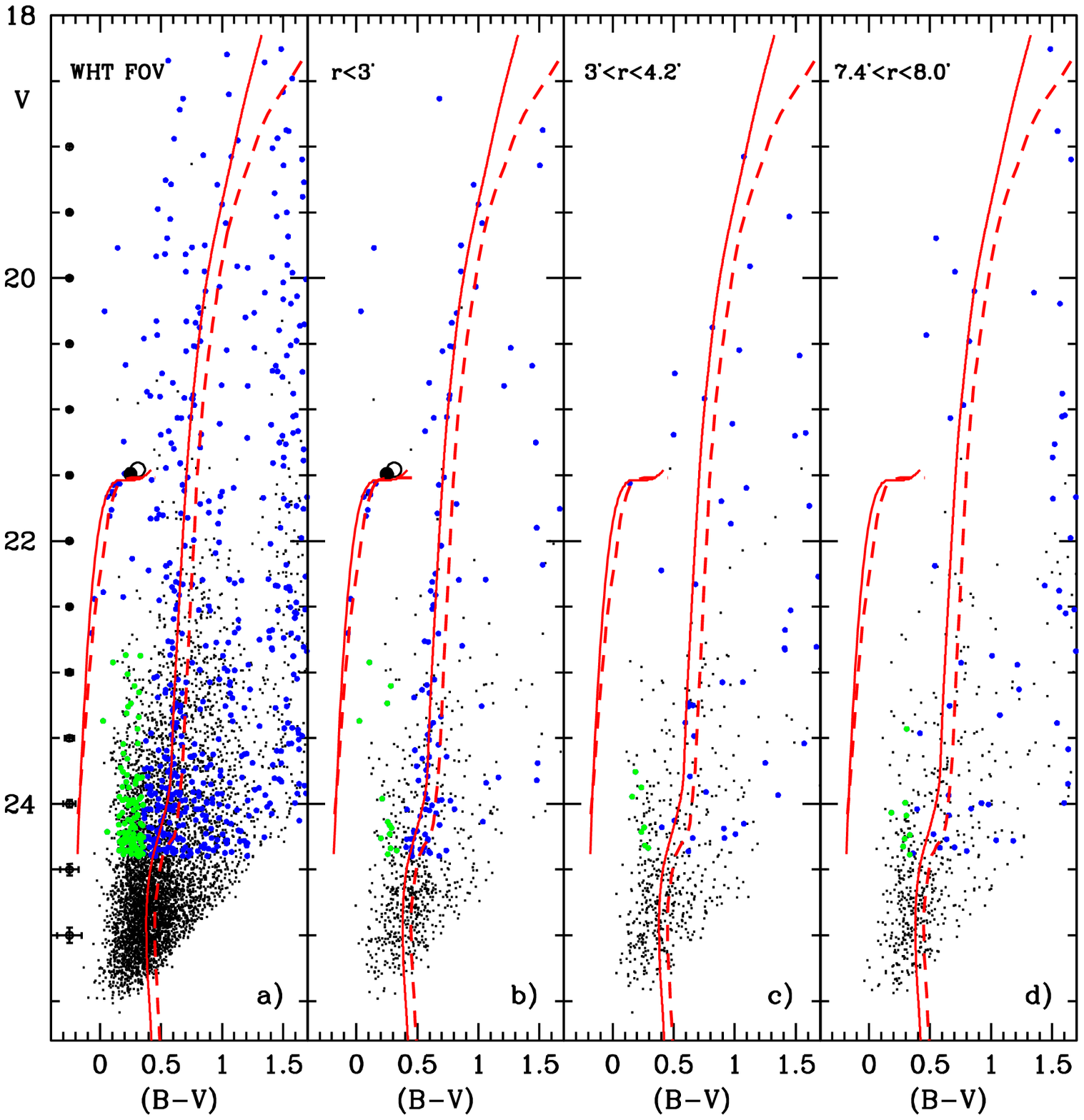}
\caption{$V,B-V$ CMDs of the CVn\,II dSph, obtained from stars in the 16.2
    $\times$ 16.2 arcmin$^2$ field covered by the WHT observations
    (panel {\it a}) and in 3 separate regions at increasing distance
    from the galaxy center (see labels).  Sources that SExtractor
    confirmed to be stars are plotted in blue, the open circle is
    the ab-type RR Lyrae star and the filled circle is the c-type RR Lyrae
    star, whereas green dots are stars with $V < 24.4$ mag in the region of the
  BSSs.
    Solid and dashed red lines are the mean ridge lines of M15 and M3,
    shifted in magnitude and adjusted in reddening to fit the galaxy's 
    horizontal and red giant branches. Typical error bars of the photometry
    are shown on the left-hand side.}
\label{f:fig3} 
\end{figure*}

\clearpage
\realfigure{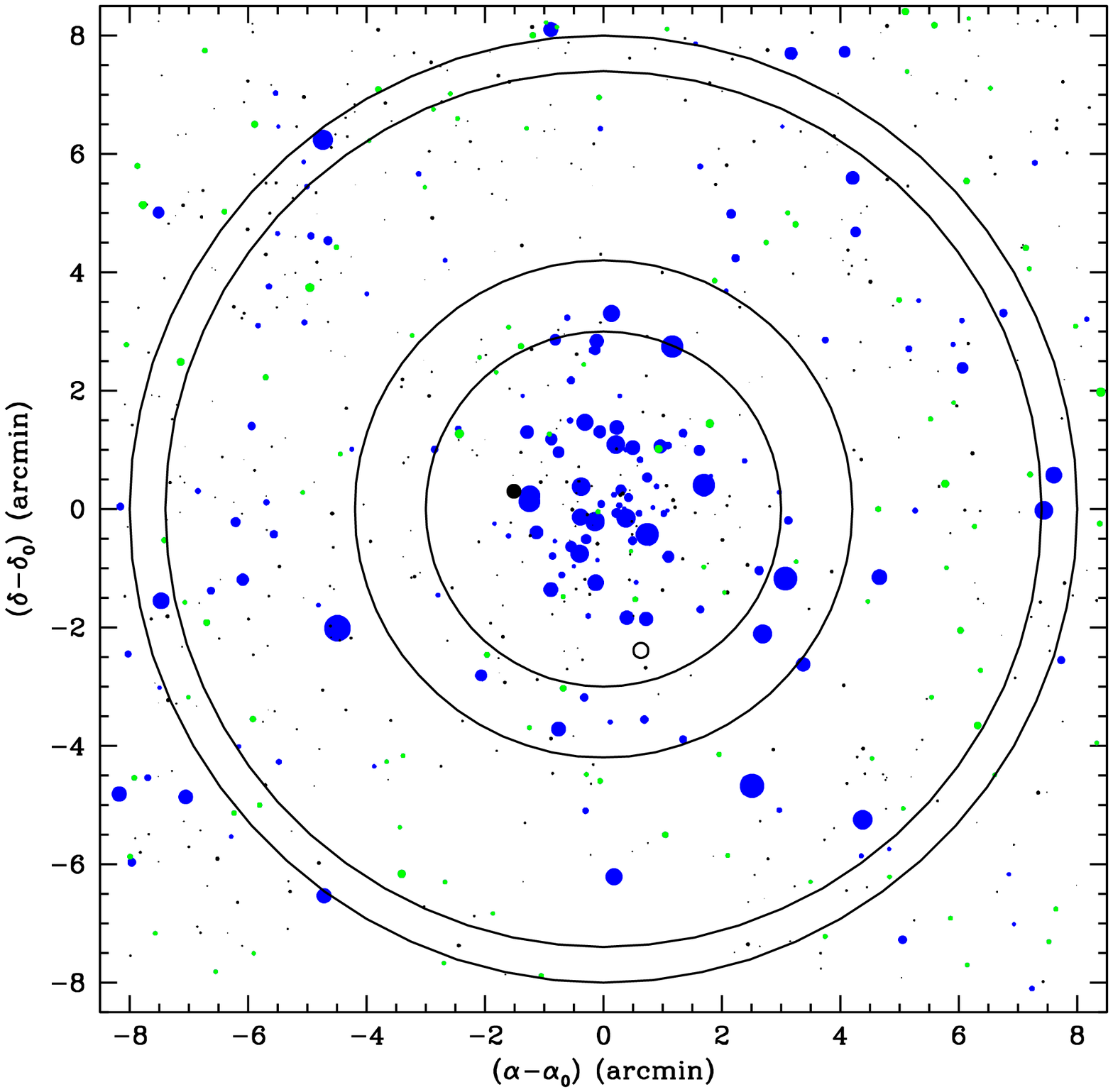}
{X,Y map, in differential R.A. and DEC (in arcmin) from the
  galaxy center, 
  set at RA(J2000)= 12$^h$ 57$^m$ 10.1$^s$,
  DEC(J2000)=34$^\circ$ 19$^\prime$ 13.5$^{\prime\prime}$,
  of sources that in Fig. 3{\it a}
  lie either within
  $\pm$ 0.1 mag from the mean ridge line of M15 or in the BSS region (green dots).
  Dot sizes are proportional to the objects' magnitudes with the
  largest dots corresponding to $V\sim$ 18-18.5 mag and the smallest ones to $V\sim$ 25 mag. 
  North is up
  and East to the left.  
  Circles have radii  of $3.0^{\prime}, 4.2^{\prime}, 7.4^{\prime}$, and $8.0^{\prime}$,
  respectively.
  Symbols and color-coding are as in
  Figure~\ref{f:fig3}. 
  }{f:fig4}

\clearpage


\begin{thebibliography}{}
\bibitem[Bellazzini(2007)]{bell07} Bellazzini, M., et al. 2007, \aap, 473, 171
\bibitem[Belokurov et al.(2006)]{be06} Belokurov, V., et al. 2006, \apj, 647, L111
\bibitem[Belokurov et al.(2007)]{be07} Belokurov, V., et al. 2007, \apj, 654, 897
\bibitem[Bertin \& Arnouts (1996)]{ber96} Bertin, A., \& Arnouts, S., 1996, \aaps, 117, 393	 
\bibitem[Buonanno et al. (1994)]{buo94} Buonanno, R., Corsi, C.E., Buzzoni, A., Cacciari, C., 
Ferraro, F.R., \& Fusi Pecci, F. 1994, \aap, 290, 69 
\bibitem[Cacciari \& Clementini(2003)]{cc03} Cacciari, C., \& Clementini, G.
2003, in 
Stellar Candles for the Extragalactic Distance Scale, ed. D. Alloin \& W.Gieren
(Berlin: 
Springer), 105
\bibitem[Carretta \& Gratton(1997)]{cg97} Carretta, E., \& Gratton, R.J. 1997,
\aaps, 121, 
95
\bibitem[Catelan(2004)]{ca04} Catelan, M. 2004, in Variable Stars in the Local 
Group, ASP.
Conf.Ser., 310, ed. D.W. Kurtz \& K.R. Pollard (San Francisco: ASP), 113  
\bibitem[Catelan(2005)]{ca05} Catelan, M. 2005, preprint (astro-ph/0507464)
\bibitem[Clement \& Rowe(2000)]{cr00} Clement, C.M., \& Rowe, J. 2000, \aj, 120,
2579
\bibitem[Clementini et al.(2000)]{clementini00} Clementini, G., et al. 
2000, \aj, 120, 2054
\bibitem[Clementini et al.(2003)]{cg03} Clementini, G., Gratton, R.G., 
Bragaglia, A., Clementini, G., Carretta, E., Di Fabrizio, L. \& Maio, M. 2003,
\aj, 125, 1309
\bibitem[Dall'Ora et al.(2006)]{dall06}Dall'Ora, M., et al. 2006, \apj, 653,
L109
\bibitem[Di Fabrizio(1999)]{df99} Di Fabrizio, L., 1999, {\emph Masters Thesis}, 
{\emph Universit\'a degli Studi di Bologna}
\bibitem[Durrell \& Harris(1993)]{dh93} Durrell, P.R., \& Harris, W.E. 1993, \aj, 105, 1420
\bibitem[Harris (1996)]{ha96} Harris, W.E. 1996, \aj, 112, 1487
\bibitem[Gratton et al.(2004)]{gratton04} Gratton, R.G., Bragaglia, A., Clementini, G., Carretta,
E., Di Fabrizio, L., Maio, M., \& Taribello, E. 2004, \aap, 412, 937
\bibitem[Grillmair(2006)]{gr06} Grillmair, C.J. 2006, \apj, 645, L37
\bibitem[Ibata et al.(1995)]{ibata95} Ibata, R.A., Gilmore, G., \& Irwin, M.J. 1995, \mnras, 277, 781 
\bibitem[Irwin et al.(2007)]{irw07} Irwin, M.J., et al. 2007, \apj, 656, L13
\bibitem[Landolt(1992)]{la92} Landolt, A.U. 1992, \aj, 104, 340
\bibitem[Kuehn et al.(2007)] {kue07} Kuehn, C., et al. 2007, ApJ (Letters), submitted
(astro-ph/0709.3281)
\bibitem[Martin et al.(2004)]{martin04} Martin, N.F., Ibata, R.A., Bellazzini, M., Irwin, M.J., Lewis, G.F., \& Dehnen, W. 2004, \mnras, 348, 12
\bibitem[Martin et al.(2007)]{martin07} Martin, N.F., Ibata, R.A., Chapman, S.C.,
Irwin, M.J.,\& Lewis, G.F. 2007b, \mnras, 380, 281
\bibitem[Mateo(1998)]{mateo98} Mateo, M.L. 1998, ARA\&A, 36, 435
\bibitem[Oosterhoff(1939)]{oo39} Oosterhoff, P. Th. 1939, Observatory, 62, 104
\bibitem[Pierce \& Nations(2002)]{p02}Pierce, M.J., \& Nations, H.L.
2002, BAAS, 34, 749
\bibitem[Rutledge et al.(1997)]{ru97} Rutledge, G.A., Hesser, J.E., \& Stetson,
P.B. 1997, 
\pasp, 109, 907
\bibitem[Schlegel et al.(1998)Schlegel, Finkbeiner, \& Davis]{sc98}Schlegel, D.J., Finkbeiner, D.P., \&
Davis, M. 1998, \apj, 500, 525
\bibitem[Schwarzenberg-Czerny(1996)]{sc96} Schwarzenberg-Czerny, A. 1996, \apj, 460, L107
\bibitem[Siegel(2006)]{s06}Siegel, M.H. 2006, \apj, 649, L83
\bibitem[Simon \& Geha(2007)]{sg07} Simon, J.D., \& Geha, M. 2007, \apj, 670, 313
\bibitem[Stetson(1987)]{st87} Stetson, P.B. 1987, \pasp, 99, 191
\bibitem[Stetson(1994)]{st94} Stetson, P.B. 1994, \pasp, 106, 250
\bibitem[van den Bergh(1995)]{vdb95} van den Bergh, S. 1995, \aj, 110, 1171; erratum: \aj, 110, 3119
\bibitem[Walsh, Jerjen, \& Willman(2007)]{wal07} Walsh, S.~M., Jerjen, H.,
\& Willman, B. 2007, \apj, 662, L83 
\bibitem[Willman et al.(2005)]{will05} Willman, B., et al. 2005a, \apj, 626, L85
\bibitem[York et al.(2000)]{yo00} York, D.G., et al. 2000, \aj, 120, 1579
\bibitem[Zinn \& West(1984)]{zw84} Zinn, R., \& West, M.J. 1984, \apjs, 55, 45
\bibitem[Zucker et al.(2006a)]{zu06a} Zucker, D.B., et al. 2006a, \apj, 643, L103
\bibitem[Zucker et al.(2006b)]{zu06b} Zucker, D.B., et al. 2006b, \apj, 650, L41

\end{thebibliography}
\end{document}